\documentclass[prl,aps,twocolumn,preprintnumbers,showpacs,superscriptaddress]{revtex4}
\usepackage{epsfig,amssymb,amsfonts}
\def\id{\mathbb{I}}
\newcommand{\ket}[1]{|#1\rangle}

\newcommand{\half}{\mbox{$\textstyle \frac{1}{2}$}}
\def\opone{\leavevmode\hbox{\small1\kern-3.8pt\normalsize1}}
\newcommand{\tr}[1]{\mbox{Tr} #1}
\newcommand{\vis}{\text{v}}

\newcommand{\depol}{\mathcal{D}}

\newcommand{\pr}{\text{P}}
\newcommand{\va}{V_{A}^{(k)}}

\newcommand{\vb}{V_{B}^{(k)}}

\newcommand{\bound}{\mathcal{B}}
\newcommand{\hilb}{\mathcal{H}}

\begin{document}
\title{Direct estimation of functionals of density operators by
local operations and classical communication}

\date{\today}
    \author{Carolina \surname{Moura Alves}}\email{carolina.mouraalves@qubit.org}%
    \affiliation{Claredon Laboratory, University of Oxford, Parks Road, Oxford OX1 3PU, U.K.}%
    \affiliation{Centre for Quantum Computation, DAMTP,
      University of Cambridge, Wilberforce Road, Cambridge CB3 0WA, U.K.}%
    \author{Pawe{\l} \surname{Horodecki}}%
    \affiliation{Faculty of Applied Physics and Mathematics, Technical
      University of Gda\'nsk, 80-952
      Gda\'nsk, Poland.} %
    \author{Daniel K. L. \surname{Oi}}%
    \affiliation{Centre for Quantum Computation, DAMTP,
      University of Cambridge, Wilberforce Road, Cambridge CB3 0WA, U.K.}%
    \author{L. C. \surname{Kwek}}%
    \affiliation{Department of Natural Sciences, National Institute of
      Education, Nanyang Technological University, 1 Nanyang Walk, Singapore
      637616}
    \author{Artur K. \surname{Ekert}}
    \affiliation{Centre for Quantum Computation, DAMTP,
      University of Cambridge, Wilberforce Road, Cambridge CB3 0WA, U.K.}%
      \affiliation{Department of Physics, National University of
    Singapore, 2 Science Drive 3, Singapore 117542.}


\begin{abstract}
  We present a method of determining important properties of a shared bipartite
  quantum state, within the ``distant labs'' paradigm, using \emph{only} local
  operations and classical communication (LOCC). We apply this procedure to
  spectrum estimation of shared states, and locally implementable structural
  physical approximations to incompletely positive maps. This procedure can
  also be applied to the estimation of channel capacity and measures of
  entanglement.
\end{abstract}

\pacs{03.67.-a,03.67.Hk,03.67.Mn}
\keywords{LOCC, structural physical approximation, functionals}

\maketitle

There are many scenarios in quantum information science where it
is necessary to estimate certain properties of a quantum state
$\varrho$, such as its spectrum, purity or degree of entanglement.
Moreover, such estimations are often needed when $\varrho$ is a
bipartite state $\varrho_{AB}$, shared by two distant parties,
Alice and Bob, who can perform only local operations and
communicate classically (LOCC). The desired properties can be then
estimated either by resorting to quantum state
tomography~\cite{Vogel1989} or more directly, e.g. via estimating
non-linear functionals of $\varrho_{AB}$. The second method has
the natural advantage of being more efficient, since we compute
directly the desired properties without estimating any superfluous
parameters. In fact the direct estimation has been successfully
applied to local spectrum estimation~\cite{direct}, entanglement
detection~\cite{direct,SPA} and the evaluation of one-qubit
quantum channel capacities~\cite{EMOHHK2002}. However, the LOCC
version of these techniques was left as an open problem. In this
paper we show that the two basic techniques, namely, the
estimation of non-linear functionals of quantum states and
constructions of Structural Physical
Approximations~\cite{direct,SPA} admit LOCC implementation. This
opens the possibility of the direct estimation of entanglement and
some channel capacities using only LOCC.

\begin{figure}
\begin{center}
\epsfig{figure=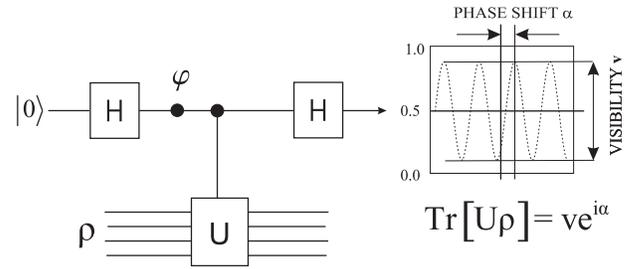,width=0.45\textwidth}
\end{center}
\caption{A modified Mach-Zender interferometer with coupling to an ancilla
  by a controlled-$U$ gate. The interference pattern is modified by the factor
  $\vis e^{i\alpha}=\tr\left[U\rho\right]$.}
\label{fig:modmachzender}
\end{figure}

As a general remark, let us recall that a quantum operation
$\Lambda$ can be implemented using LOCC if it can be written as a
convex sum
\begin{equation}
\Lambda = \sum_k p_k \; A_k\otimes B_k, \label{eq:locc}
\end{equation}
where $A_k$ acts on the subsystem at Alice's location and $B_k$ on
the subsystem at Bob's location, and $p_k$ represent the
respective probabilities.


Let us start with the estimation of non-linear functionals of
$\varrho_{AB}$ using quantum interferometry. Consider a typical
set-up for single qubit interferometry, conveniently expressed in
terms of quantum gates and networks: Hadamard gate, phase-shift
gate, Hadamard gate, and measurement in the computational basis
$\{\ket{0},\ket{1}\}$. We modify the interferometer by inserting a
controlled-$U$ operation between the Hadamard gates, with its
control on the qubit and with $U$ acting on a quantum state
$\rho$~(Fig.~\ref{fig:modmachzender}). The controlled-$U$ models
the interaction between the qubit and an auxiliary system
(ancilla), initially in the state $\rho$, and it leads to
modification of the observed interference pattern, by the factor
$\vis e^{i\alpha}=\tr\left[U\rho\right]$. The factor $\vis$ is the
new visibility and $\alpha$ is the shift of the interference
fringes, known as the Pancharatnam phase~\cite{Pancha56}. The
observed modification of the fringes gives us an estimate of the
average value of unitary operator $U$ in state
$\rho$~\cite{sjoqvist2000}.

Suppose now that $\rho$ is the quantum state of two separable
subsystems, $\rho=\varrho_A\otimes\varrho_B$ and that we choose
$U$ to be the swap operator $V$, defined such that
$V\ket{\phi}_A\ket{\psi}_B=\ket{\psi}_A\ket{\phi}_B$,
$\forall\ket{\phi},\ket{\psi}$. In this case, the modification of
the interference pattern will be $\vis=\tr\left[V(\varrho_A\otimes
\varrho_B)\right]=\tr\left[\varrho_A\varrho_B\right]$, or the
overlap between the input states $\varrho_A$ and $\varrho_B$. If
the two inputs states are equal, $\varrho_A=\varrho_B=\varrho$, we
obtain an estimation of the purity, $\tr\left[\varrho^2\right]$.
The generalization of the swap operation $V$ to the shift
operation $V^{(k)}$ ($V^{(k)}\ket{\phi_1}\ket{\phi_2}...
\ket{\phi_k}=\ket{\phi_k}\ket{\phi_1}...\ket{\phi_{k-1}}$,
$\forall\ket{\phi_i}$, $i=1,...,k$), and the choice of
$\rho=\varrho^{\otimes{k}}$ as the input state, allows us to
estimate multi-copy observables, $\tr[\varrho^k]$, of an unknown
state $\varrho$~\cite{EMOHHK2002,direct,SPA}.

\begin{figure}
\begin{center}
\epsfig{figure=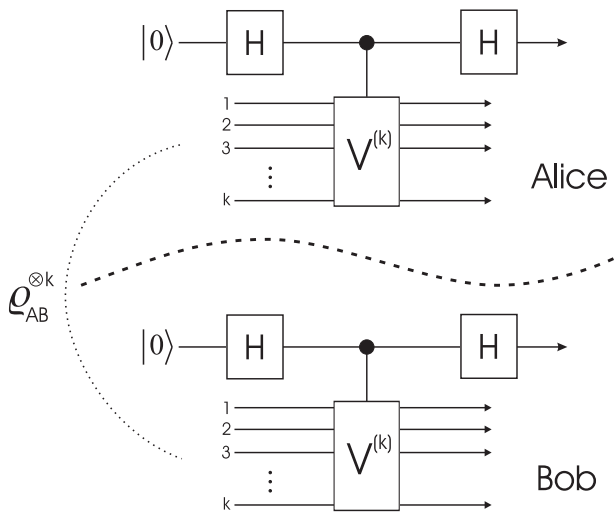,width=0.45\textwidth}
\end{center}
\caption{Network for remote estimation of non-linear functionals
of bipartite density operators. Since $\tr[V^{(k)}\varrho^{\otimes
k}]$ is real, Alice and Bob can omit their respective phase
shifters.} \label{fig:remoteest}
\end{figure}

Let us now extend this method to the LOCC scenario by constructing
two local networks, one for Alice and one for Bob, in such a way
that the global network is similar to the network with the
controlled-shift. Unfortunately, the global shift operation
$V^{(k)}$ cannot be implemented using only LOCC, since it does not
admit decomposition~(\ref{eq:locc}). Thus, we will implement it
indirectly, using the global network shown in
Fig.~\ref{fig:remoteest}. Alice and Bob share a number of copies
of the state $\varrho_{AB}\in\bound(\hilb^d)$. They group them
respectively into sets of $k$ elements, and run the local
interferometric network on their respective halves of the state
$\rho_{AB}=\varrho_{AB}^{\otimes k}$. For each run of the
experiment, they record and communicate their result.

The individual interference patterns Alice and Bob record will
depend only on their respective reduced density operators. Alice
will observe the visibility $\vis_A=\tr[\varrho_{A}^{k}]$ and Bob
will observe the visibility $\vis_B=\tr[\varrho_{B}^{k}]$.
However, if they compare their individual observations, they will
be able to extract information about the global density operator
$\varrho_{AB}$, e.g. about
\begin{equation}
\tr[\varrho_{AB}^k]=\tr\left[\varrho_{AB}^{\otimes
k}\;\left(\va\otimes\vb\right)\right].
\end{equation}
This is because Alice and Bob can estimate the probabilities
$\pr_{ij}$ that in the measurement Alice's interfering qubit is
found in state $\ket{i}_A$ and Bob's in state $\ket{j}_A$ for $i,j
= 0,1$. These probabilities can be conveniently expressed as
\begin{small}
\begin{equation}
\pr_{ij}=\frac{1}{4}\tr\Big[\varrho_{AB}^{\otimes k}\; \big(
\id+(-1)^i \va\big) \otimes \big(\id + (-1)^j \vb\big)\Big],
\label{eq:probs}
\end{equation}
\end{small}
hence the formula for the basic non-linear functional of
$\varrho_{AB}$ reads
\begin{equation}
\tr[\varrho_{AB}^k] = \pr_{00}-\pr_{01}-\pr_{10}+\pr_{11}.
\label{eq:sigmazsigmaz}
\end{equation}
In fact, the expression above is the expectation value
$\langle\sigma_{z}\otimes\sigma_{z}\rangle$, measured on Alice's
and Bob's qubits (the two qubits that undergo interference). Given
that we are able to directly estimate $\tr[\varrho_{AB}^k]$ for
any integer value of $k$, we can estimate the spectrum of
$\varrho_{AB}$ without resorting to a full state tomography.


We next show how to implement Structural Physical Approximations
within the LOCC constraint. Structural Physical Approximations
(SPAs) were introduced recently as tools for determining relevant
parameters of density operators (see~\cite{SPA,direct} for more
details). Basically the SPA of a mathematical operation $\Lambda$,
denoted as $\tilde\Lambda$, is a physical operation, a process
that can be carried out in a laboratory, that emulates the
character of $\Lambda$. More precisely, suppose
$\Lambda:\bound(\hilb^d)\mapsto\bound(\hilb^d)$ is a trace
preserving map which does not represent any physical process, for
example, an anti-unitary operation such as transposition. Then a
convex sum
\begin{equation}
\tilde\Lambda = \alpha\depol+(1-\alpha)\Lambda, \label{eq:SPA}
\end{equation}
where $\depol$ is the depolarizing map which sends any density
operator into the maximally mixed state, represents a physical
process, i.e. a completely positive map, as long as $\alpha$ is
sufficiently large. On top of this $\depol$, with its trivial
structure, does not mask the structure of $\Lambda$. The
Structural Physical Approximation to $\Lambda$ is obtained by
selecting, in the expression above, the threshold value
$\alpha=(d^{2}\lambda)/(d^{2}\lambda+1)$, where $-\lambda$ is the
lowest eigenvalue of $(\id\otimes\Lambda)P_+^{(d)}$ and
$P_+^{(d)}$ is a maximally entangled state of a $d\times d$
system~\footnote{The threshold value for $\alpha$ is obtained from
the requirement of complete positivity of $\tilde\Lambda$, which
in this case can be reduced to $\tilde\Lambda P_+^{(d)}\ge 0$}.

Please note that the physical implementation of SPAs is not a
trivial problem as the formula~(\ref{eq:SPA}), which explicitly
contains the physically impossible map $\Lambda$, is of little
guidance here. Let us also mention in passing that if $\Lambda$ is
not trace preserving then $\tilde\Lambda$ may be implementable but
only in a probabilistic sense e.g. via a post-selection.

There are many examples of mathematical operations which although
important in the formulation of the physical theory do not
represent a physical process. For example, mathematical criteria
for entanglement involve positive but not completely positive
maps~\cite{Horodeckis1996} and as such they are not directly
implementable in a laboratory --- they tacitly assume that a
precise description of a quantum state of a physical system is
given and that such operations are performed on the mathematical
description of the state rather than the system itself.

If $\Lambda$ does not represent any physical process then its
trivial extension to a bipartite case, $\id\otimes\Lambda$, does
not represent a physical process either. Still, its SPA,
$\widetilde{\id\otimes\Lambda}$, does describe a physical
operation, but can it be implemented with LOCC?

The positive answer is obtained by putting
$\widetilde{\id\otimes\Lambda}$ into the tensor product
form~(\ref{eq:locc}). Let us start by writing it as
\begin{small}
\begin{eqnarray}
\widetilde{\id\otimes\Lambda}
&=&\alpha\depol\otimes\depol+(1- \alpha)\id\otimes\Lambda\nonumber\\
&=&(1-\alpha+\beta)\id\otimes
\left(\frac{1-\alpha}{1-\alpha+\beta}\Lambda +\frac{\beta}{1-\alpha+\beta}\depol\right)\nonumber\\
&+&(\alpha-\beta)\left(\frac{\alpha}{\alpha-\beta}\depol+
\frac{-\beta}{\alpha-\beta}\id\right)\otimes\depol\nonumber\\
&=&(1-\alpha+\beta)\id\otimes\tilde{\Lambda}+(\alpha-\beta)
\tilde{\Theta}\otimes\depol, \label{eq:remotespa}
\end{eqnarray}
\end{small}
where
\begin{eqnarray}
\tilde{\Lambda}&=&\frac{1-\alpha}{1-\alpha+\beta}\Lambda
+\frac{\beta}{1-\alpha+\beta}\depol,\\
\tilde{\Theta}&=&\frac{\alpha}{\alpha-\beta}\depol+
\frac{\beta}{\alpha-\beta}(-\id). \label{mapstwo}
\end{eqnarray}
Equation~(\ref{eq:remotespa}) does not represent a convex sum of
physically implementable maps for any values of $\alpha$ and
$\beta$ but if we choose
\begin{eqnarray}
\beta\ge(1-\alpha) \lambda d^{2}\label{eq:ineq1}\\
\alpha\ge \beta d^{2}, \label{eq:ineq2}
\end{eqnarray}
where $-\lambda$ is the minimum eigenvalue of
$\id\otimes\Lambda(P^{d}_+)$, then indeed
$\widetilde{\id\otimes\Lambda}$ is a physical operation in the
LOCC form. Note, however, that the map $\tilde{\Theta}$ is not
trace preserving and as such it can be implemented only with a
certain probability of success. The minimal parameters $\alpha$
and $\beta$ that satisfy inequalities Eqs.~(\ref{eq:ineq1})
and(\ref{eq:ineq2}) are
\begin{eqnarray}
\alpha = \frac{\lambda d^4}{\lambda d^4 +1},\label{eq:alpha*}\\
\beta = \frac{\lambda d^{2}}{\lambda d^4 +1}.\label{eq:beta*}
\end{eqnarray}
Hence, the SPA $\widetilde{\id\otimes\Lambda}$ can be implemented,
by Alice and Bob, using only only LOCC.

The SPAs have been employed to test for quantum
entanglement~\cite{direct}. Recall that a necessary and sufficient
condition for a bi-partite state $\varrho_{AB}$ to be separable is
$\id\otimes\Lambda(\varrho_{AB})\ge 0$, for all positive maps
$\Lambda$~\cite{Horodeckis1996}. This condition, when considering
the SPA $\widetilde{\id\otimes\Lambda}$ on $\varrho_{AB}$, is
equivalent to
\begin{equation}
\left[\widetilde{\id\otimes\Lambda}\right]\varrho_{AB}\ge
\frac{d^{2}\lambda}{d^{4}\lambda+1},
\end{equation}
where $-\lambda$ is the minimal eigenvalue of the state
$\left[(\id\otimes\id)\otimes(\id\otimes
\Lambda)\right](P^{d^2}_+)$~\cite{direct}. Thus, by estimating the
spectrum (or the lowest eigenvalue) of the state
$\left[\widetilde{\id\otimes\Lambda}\right]\varrho_{AB}$, we can
directly detect quantum entanglement. Moreover, we have already
shown that both $\widetilde{\id\otimes\Lambda}$ and the spectrum
estimation of $\varrho_{AB}$, via non-linear functionals, can be
performed using only LOCC, hence a direct detection of quantum
entanglement within the LOCC scenario is also possible.


Let us now comment briefly on other potential applications of the
methods presented above. Let a completely positive map
$\Lambda:\bound (\hilb^d)\mapsto\bound(\hilb^d)$ represent a
quantum channel shared by Alice and Bob. An estimation of the
channel capacity may involve either a channel tomography or a
direct estimation. In the case of tomography Alice prepares a
maximally entangled pair of particles in state $P_+^{d}$ and sends
one half of the pair to Bob. They now share the state
\begin{equation}
\varrho_\Lambda=\left[\id\otimes\Lambda\right] P_+^d.
\end{equation}
From the Jamio{\l}kowski isomorphism~\cite{Jamiolkowski1972}, this
bi-partite state encodes all properties of the channel $\Lambda$,
so state tomography on $\varrho_\Lambda$ is effectively channel
tomography on $\Lambda$. However, given a bi-partite state
$\varrho_\Lambda$, Alice and Bob can also use the LOCC techniques
to directly estimate its desired properties. For example, it has
been shown that a single qubit channel $\Lambda$ has non-zero
channel capacity if and only if the maximal eigenvalue of
$\varrho_\Lambda$ is strictly greater than $\half$
(see~\cite{EMOHHK2002} for details). This can be estimated
directly via the spectrum estimation, which in the case of two
qubits requires only $2\times 4 -3=5$ measurements of the type
$\sigma_z\otimes\sigma_z$ as opposed to the $15$ parameters
required for the state estimation.

For Bell diagonal states (i.e. two-qubit states, whose
eigenvectors are all maximally entangled), the entanglement of
formation (or negativity, see below) can be inferred from its
spectrum~\cite{hugepaper}. Thus, if Alice and Bob share a Bell
entangled state, they can estimate the degree of entanglement of
their state through spectrum estimation only. An important
subclass of Bell diagonal states are the maximally correlated
states, rank two states equivalent (up to $U_A\otimes U_B$
transformations) to mixtures of two pure states,
$\ket{\psi_+}=\frac{1}{\sqrt{2}}(\ket{0}\ket{0}+\ket{1}\ket{1})$
and $\ket{\psi_-}=\frac{1}{\sqrt{2}}
(\ket{0}\ket{0}-\ket{1}\ket{1})$. The one-way distillable
entanglement can be calculated for such states as
$D_\rightarrow=\log 2 -S(\varrho)$, which is a function solely of
the spectrum. Thus, instead of estimating the seven parameters
required to describe maximally correlated states, we need only
estimate five parameters.

The estimation of entanglement measures (see~\cite{QIC} for
review) is known  only for special cases, such as the computable
measure of entanglement~\cite{WernerVidal},
 $\mathcal{N}(\varrho_{AB})\equiv\log||\varrho_{AB}^{T_B}||=\log(\sum_i
|\lambda_i|)$.  This measure is valid for any shared bipartite
state, with a maximally mixed reduced density operator of at least
one sub-system, and it is a function of the spectrum
$\{\lambda_i\}$ of the partially transposed matrix
$\varrho_{AB}^{T_B}\equiv\id\otimes T(\varrho_{AB})$, where $T$ is
the (incompletely positive) transposition map. Thus, we can
estimate $\mathcal{N}(\varrho_{AB})$ using only LOCC, if we choose
$\Lambda=T$ and then estimate the spectrum of the resultant state.

Given any quantum channel $\Lambda$, $\mathcal{N}(
\varrho_\Lambda)$ is the upper bound for one-way channel capacity.
We obtain, therefore, a necessary condition for non-zero one-way
capacity $Q_\rightarrow$: if $\mathcal{N}(\varrho_\Lambda)=0$, the
two-way channel capacity must \emph{vanish}~\cite{HolevoWerner}
(this can be easily seen using distillation and binding
entanglement channel~\cite{Bechan}). Hence, the positive partial
transpose (PPT) test is a strong necessary test of quantum
non-zero channel capacity.

To conclude we have demonstrated that both direct spectrum
estimations and the structural physical approximations can be
implemented in the case of bi-partite states using only local
operations and classical communication. This leads to more direct,
LOCC type,  detections and estimations of quantum entanglement and
of some properties of quantum channels. Direct estimations of
specific properties have the natural advantage over the state
tomography because they avoid estimating superfluous parameters.
Still, the exact comparison of the use of physical resources in
tomography and direct estimations depends very much on the
physical implementations of these techniques. Our objective here
is to provide additional tools for quantum information processing
rather than comparative studies of these tools.

A.K.E. and L.C.K. acknowledge financial support provided under the
A$^*$Star Grant No.\ 012-104-0040. P.H. acknowledges support from
the Polish Committee for Scientific Research and the European
Commission. C.M.A. is supported by the Funda{\c c}{\~a}o para a
Ci{\^e}ncia e Tecnologia (Portugal) and D.K.L.O would like to
acknowledge the support of the Cambridge-MIT Institute Quantum
Information Initiative, and EU projects RESQ (IST-2001-37559) and
TOPQIP (IST-2001-39215).

\end{document}